# Identification of the vortex glass phase by harmonics of the AC magnetic susceptibility

M. G. Adesso[*], M. Polichetti, S. Pace

*Dipartimento di Fisica, Universita' degli Studi di Salerno & INFM-SUPERMAT*

*Via S. Allende, 84081, Baronissi (SALERNO) - ITALY*

**Abstract**

We compared the AC magnetic susceptibility behaviour for the vortex glass phase and for the creep phenomena with an inhomogeneous pinning potential. The temperature dependence of the harmonics of the susceptibility have been numerically simulated with these two models, and we studied them as a function of the frequency, in terms of Cole-Cole plots. From our analysis we show that it is possible to distinguish between the two different phases, because of their clear differences in the Cole-Cole plots behaviour with the frequency.

---

[*] Corresponding author. Tel.: ++39-89-965 256; fax: ++39-89-965 275; e-mail: adesso@sa.infn.it.

## 1. Introduction

The study of vortex dynamics in type II superconductors is a complex topic and many open questions are still present [1]. In particular, the existence of a real vortex glass phase in the mixed state of type-II superconductors [1] is argument of discussion. Koch. et al. [2] showed that some Voltage-Current characteristics (V-I) in YBCO film samples, can be interpreted in terms of a glass phase, characterised by a non linear U(J) dependence of the pinning potential on the current density J. Different creep models have been developed in order to take into account this phase [1,3-5], as alternative to the Kim-Anderson model [6], in which U(J) is linear. Among them the most used is the *vortex glass collective creep* model [3-5].

Nevertheless, the interpretation of the Koch experimental data [2] is very controversial: Coppersmith [7] and Landau [8] showed that the standard Kim-Anderson approach, modified with a pinning potential U(*x*) depending on the spatial position *x*, also reproduces the qualitative features of the Koch's V-I characteristics [2]. Aim of this work is to verify if the vortex glass collective creep is equivalent to an inhomogeneous Kim-Anderson approach, by analysing the AC magnetic susceptibility behaviour in both these models.

## 2. Numerical technique

In order to simulate the harmonics of the AC magnetic susceptibility, we numerically integrated the non linear diffusion equation for the magnetic field inside the sample:

$$\frac{\partial B}{\partial t} = \frac{\partial}{\partial x}\left[\left(\frac{\rho(B,J)}{\mu_0}\right) \cdot \frac{\partial B}{\partial x}\right] \qquad (1)$$

in which the resistivity is:

$$\rho(B,J) = \rho_{Flow}(B) \cdot e^{-\left(\frac{U(J,x,T)}{k_B T}\right)} \qquad (2)$$

where $\rho_{Flow}$ is the Flux Flow resistivity and $k_B$ is the Boltzmann constant.

The temperature dependence of the pinning potential is related to the microscopic pinning mechanisms; in our simulations we choose the δl type collective pinning model [1,9], for which:

$$U(T) \propto \left[1 - \left(\frac{T}{T_c}\right)^4\right] \equiv f(T), \qquad (3)$$

where $T_c$ is the critical temperature.

On the contrary, the dependence on the current is associated to the different creep models. In particular, if we choose the following [10]:

$$U(J) \propto \frac{1}{\mu} \cdot \left[\left(\frac{J}{J_c(T)}\right)^\mu - 1\right] \equiv g_\mu(J), \qquad (4)$$

we can change the $\mu$ parameter to analyse both the linear and the non-linear cases. The linear case corresponds to the standard Kim-Anderson creep ($\mu = -1$), whereas nonlinearities are associated to the existence of the vortex glass phase, at the different vortex regimes ($\mu = 1/7$ for the single vortex, $\mu = 3/2$ for the small bundle and $\mu = 7/9$ for the large bundle [1]). In the (4), also the temperature dependence of the critical current density, $J_c(T)$, is chosen according to the δl type collective pinning model [1].

Finally, the dependence U(*x*) on the spatial variable is associated to an inhomogeneous distribution of the pinning centres. The choice of this dependence does not influence the main features of the analysis [8] but, as suggested by Landau, in order to reproduce the Koch [2] V-I characteristics it is necessary to include also a further temperature dependence, strictly connected to the *x*-dependence. According to Ref. [8], we choose the following model:

$$U(x,T) \propto \left( |x| - a\left[1 - \left(\frac{T}{T_c}\right)\right]^k \cdot x^2 \right) \equiv h(x,T) \quad . \tag{5}$$

Combining the (3), (4) and (5), the general expression of the pinning potential is:

$$U(J,x,T) = U_0 \cdot f(T) \cdot g_\mu(J) \cdot h(x,T) \tag{6}$$

where $U_0$ is the value of the pinning potential at zero field, temperature and current.

In particular, we simulated the harmonics of the AC susceptibility in different situations, namely:

I. homogeneous Kim-Anderson model, corresponding to: $\mu = -1$ and h(*x*,T) = constant;

II. homogeneous vortex glass – collective creep model, corresponding to $\mu = 1/7$; $\mu = 3/2$; $\mu = 7/9$; h(*x*,T) = constant;

III. inhomogeneous Kim-Anderson model, for which $\mu = -1$ and h(*x*,T) is given by the expression (5).

Parameters used in the simulations are: $U_0/k_B = 16 \cdot 10^3$ K, $T_c = 91.6$ K, $k = 1.5$, $a = 1$.

## 3. Numerical results and discussion

The analysis of the AC field amplitude ($h_{AC}$) dependence of the Cole-Cole plots has been previously performed for both the Kim-Anderson and the vortex glass creep models [11], in the homogeneous case, showing that the use of $h_{AC}$ does not allow us to distinguish between the two models. In this work, we take into account the frequency dependence of the AC response. In fact, in Fig.1 the 1st harmonics Cole-Cole plots are shown, as simulated by using models I. and II. with an homogeneous pinning potential, at different frequencies.

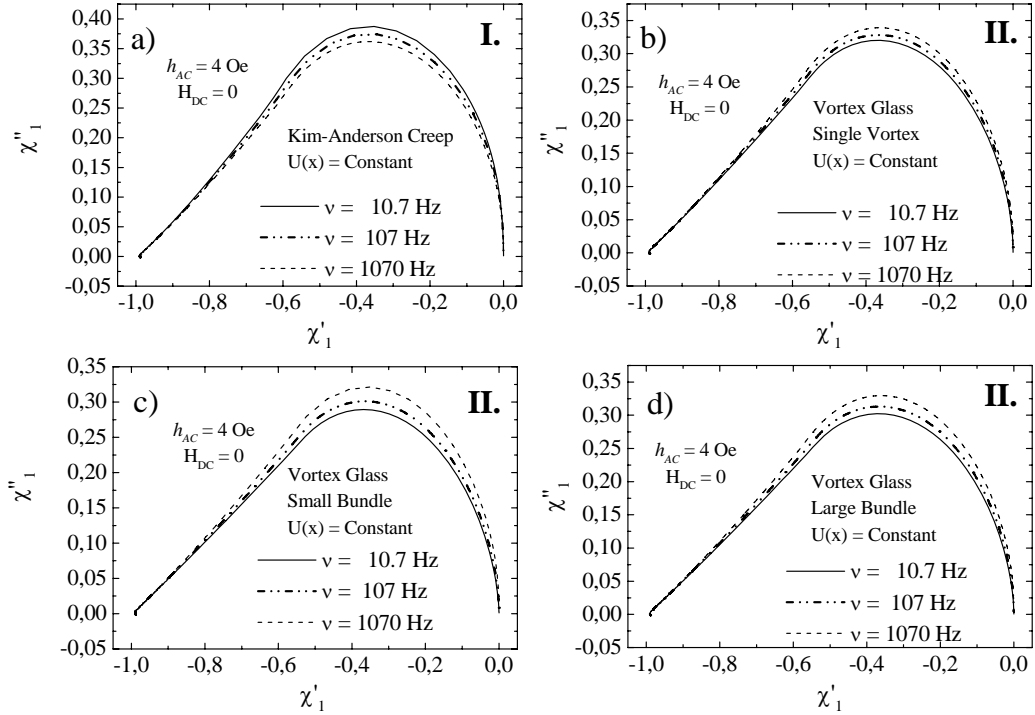

Fig.1: 1st harmonics Cole-Cole plots of the AC magnetic susceptibility, at different frequencies, simulated by using, for (a), the homogeneous Kim-Anderson creep model and, for (b), (c) and (d), the homogeneous Vortex Glass - Collective creep model, respectively in the Single Vortex (b), Small Bundle (c), and Large Bundle (d) regimes. Note that the height of the maximum, as a function of the frequency, has a different behaviour depending on if U(J) is linear (a), or non linear (b)-(d).

From this figure we can observe that the general behaviour of the plots is completely different, depending on $g_\mu(J)$. In particular, in the homogeneous Kim-Anderson model (I.), the height of the maximum in the plots decreases if the frequency is increased; on the contrary, it increases with the frequency when the vortex glass phase (II.) is considered, in all the dynamical regimes (single vortex, small bundle and large bundle). From these considerations, we can deduce that this simple analysis of the 1$^{st}$ harmonics Cole-Cole plots as a function of the frequency represents a good tool to distinguish between a linear and a non linear dependence of the pinning potential on the current, in the homogeneous case.

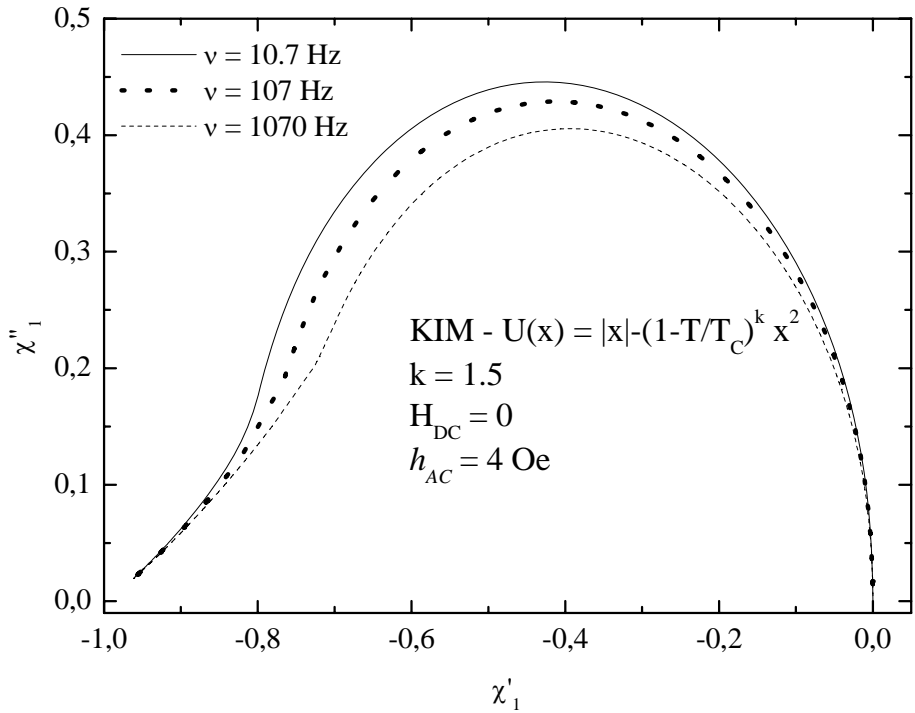

Fig.2: 1$^{st}$ harmonics Cole-Cole plots of the AC susceptibility, at different frequencies, numerically obtained by considering a pinning potential which depends on the spatial variable and, linearly, on the current (Kim-Anderson model). The maximum in the plots decreases if the frequency is increased, as in the standard Kim-Anderson model, contrarily to what happens in a Vortex Glass phase.

The frequency dependence of the 1$^{st}$ harmonics Cole-Cole plots in the inhomogeneous Kim-Anderson approach (III.) is reported in Fig.2. It is worthy of note that the maximum in the 1$^{st}$ harmonics Cole-Cole plots decreases when the frequency is increased. This behaviour is similar to what happens in the homogeneous Kim-Anderson case, whereas it is in contrast with the frequency dependence within the vortex glass models.

From this analysis, we can deduce that a standard creep model modified with a spatially inhomogeneous pinning potential, although reproducing the same V-I characteristics [8] as for the vortex glass phase, presents a different AC magnetic response respect to the latter.

**4. Conclusions**

We simulated the harmonics of the AC magnetic susceptibility by considering different dissipative phenomena. In particular, we took into account the standard Kim-Anderson Creep and the Vortex Glass Collective Creep, both with an homogeneous pinning potential, and the Kim-Anderson Creep with a pinning potential depending on the spatial variable, $x$. We analysed the obtained curves in terms of the Cole-Cole plots. In the homogeneous case, for changing AC frequencies, we observed that the $1^{st}$ harmonics Cole-Cole plots for the standard creep behave in opposite way with respect to the vortex glass case. Nevertheless, if we introduce an inhomogeneous pinning potential, the general frequency behaviour of the Cole-Cole plots for the Kim-Anderson creep remains unchanged.

We conclude that the introduction of a spatial dependence in the pinning potential is not sufficient for reproducing the glassy magnetic response, although it produces non-linearity in the Voltage-Current characteristics, which appears similar to that generated in the vortex glass phase. Therefore, this kind of frequency analysis of the AC susceptibility Cole-Cole plots can help in the detection of a vortex glass phase in type two superconductors.